\documentclass{iau}
\usepackage{graphicx}

\title[Optical observation of SN 2012A] 
{The optical photometric and spectroscopic investigation of Type IIP supernova 2012A}

\author[Roy et al.]   
{Rupak Roy$^1$, Firoza Sutaria$^2$, Subhash Bose$^1$, Sean Johnson$^3$,
 Vikram Dwarkadas$^3$, Brian York$^4$, Brijesh Kumar$^1$, Brajesh Kumar$^{1,5}$,
 Vijay K. Bhatt$^1$, Sayan Chakraborti$^6$, Don York$^3$, Adam Ritchey$^7$,
 Gabrielle Saurage$^8$ \and Mary Beth Kaiser$^9$}

\affiliation{$^1$Aryabhatta Research Institute of Observational Sciences, Nainital, India\\[\affilskip]
$^2$Indian Institute of Astrophysics, Bangalore, India\\[\affilskip]$^3$University of Chicago, Chicago, USA\\[\affilskip]$^4$Space Telescope Science Institute, Baltimore, USA\\[\affilskip]$^5$Institut d'Astrophysique et de G\'{e}ophysique, Universit\'{e} de Li\`{e}ge, Li\`{e}ge, Belgium\\[\affilskip]$^6$Institute for Theory and Computation, Harvard, USA\\[\affilskip]$^7$University of Washington, Seattle, USA\\[\affilskip]$^8$Apache Point Observatory, Sacramento Mountains, Sunspot, USA\\[\affilskip]$^9$Johns Hopkins University, Baltimore, Maryland, USA}

\pubyear{2013}
\volume{296}  
\pagerange{--}
\jname{Supernova Environmental Impacts}
\editors{Alak Ray \& Dick McCray}
\begin{document}

\maketitle

\begin{abstract}
 Supernova 2012A was discovered on 7.39UT, January, 2012 in the nearby
 galaxy NGC 3239 at an unfiltered magnitude of 14.6 and classified
 spectroscopically as a Type IIP event. Here, we present the optical
 photometric and spectroscopic follow-up of the event during 14d to 130d post
 explosion.

\end{abstract}

\firstsection 
\section{Introduction}
 Given the considerable development in automated sky survey programs,
 the detection rate of supernovae (SNe) has increased enormously
 (\cite[Lennarz et al. 2012]{Lennarz2012}) in the last few years. In a
 volumetric study it was found that more than 50\% of all SNe are
 Hydrogen rich (Type II), of which about 70\% show prolonged plateau
 (IIP) in their light curves (\cite[Li et
   al. 2011]{Li2011}). Type II SNe show diversities in their light curves and
 spectra.  Indeed there are distinct classes $-$ normal and subluminous events
 - which show different properties, along with several peculiar events like SN
 1987A, which resulted from explosion of a blue supergiant star.  Following the
 discovery of SN 2008in (\cite[Roy et
   al. 2011a]{Roy2011}), a different class of SNe has been suggested
 with characteristics in-between normal and subluminous Type IIP.  SN
 2009js is a new entry in this category (\cite[Gandhi et
   al. 2013]{Gandhi2013}). The nearby event SN 2012A also showed
 similar behaviour, and is the prime target of our study.

 SN 2012A was discovered on 7.39UT, January, 2012 in the nearby ($\sim$ 9 Mpc)
 galaxy NGC 3239 at an unfiltered magnitude of 14.6 (\cite[Moore et
   al. 2012]{Moore2012}). It was classified as a Type IIP event, with
 a spectrum similar to that of SN 2004et at about 14d post explosion,
 confirming its identity as a young Type IIP
 (\cite[Stanishev and Pursimo 2012; Roy and Chakraborti 2012]{Stanishev2012}).

\section{Observations and data reductions}
\begin{figure*}
\centering
\includegraphics[width=5.9cm]{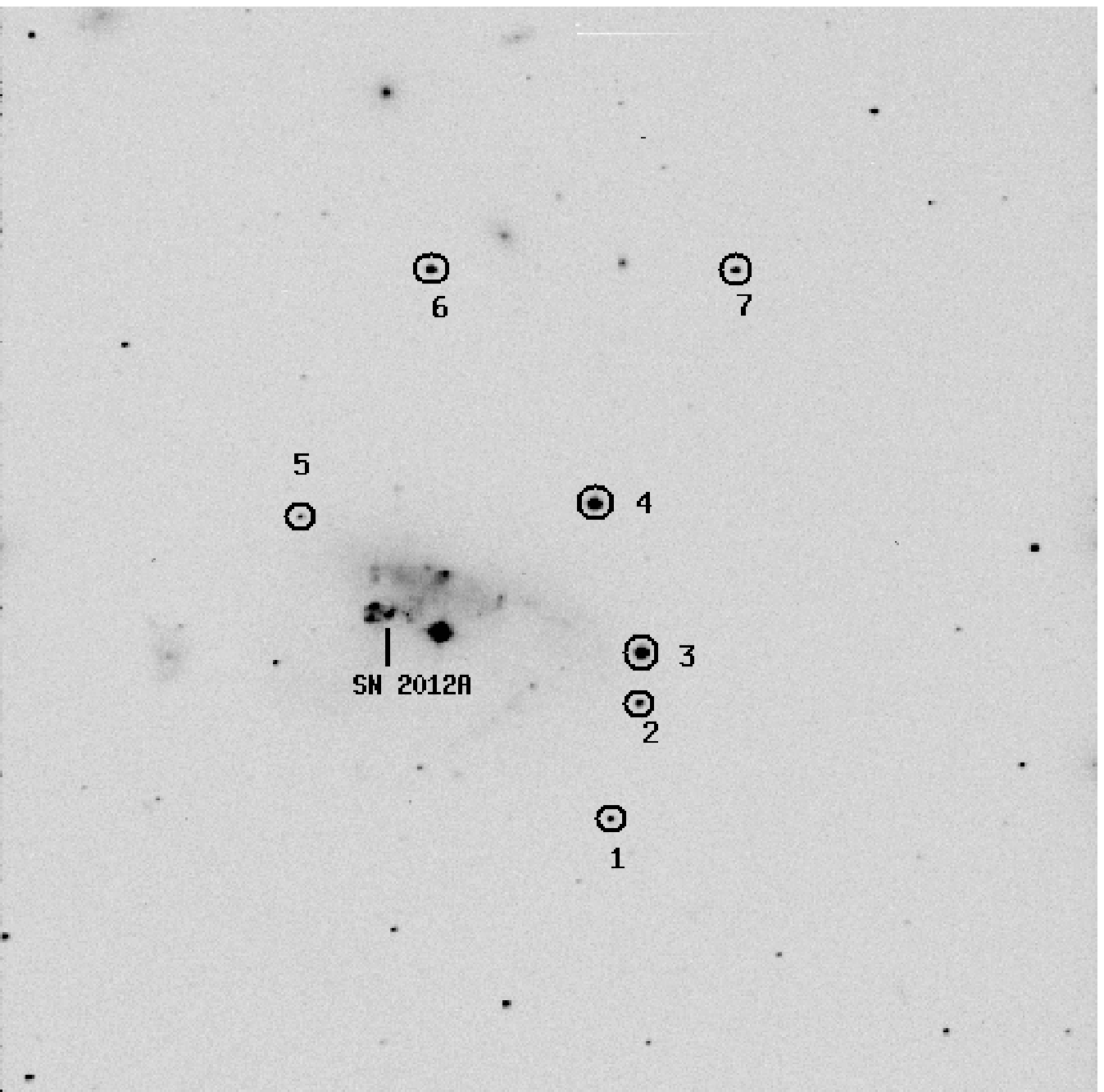}%
\hskip 2mm
\includegraphics[width=6.5cm]{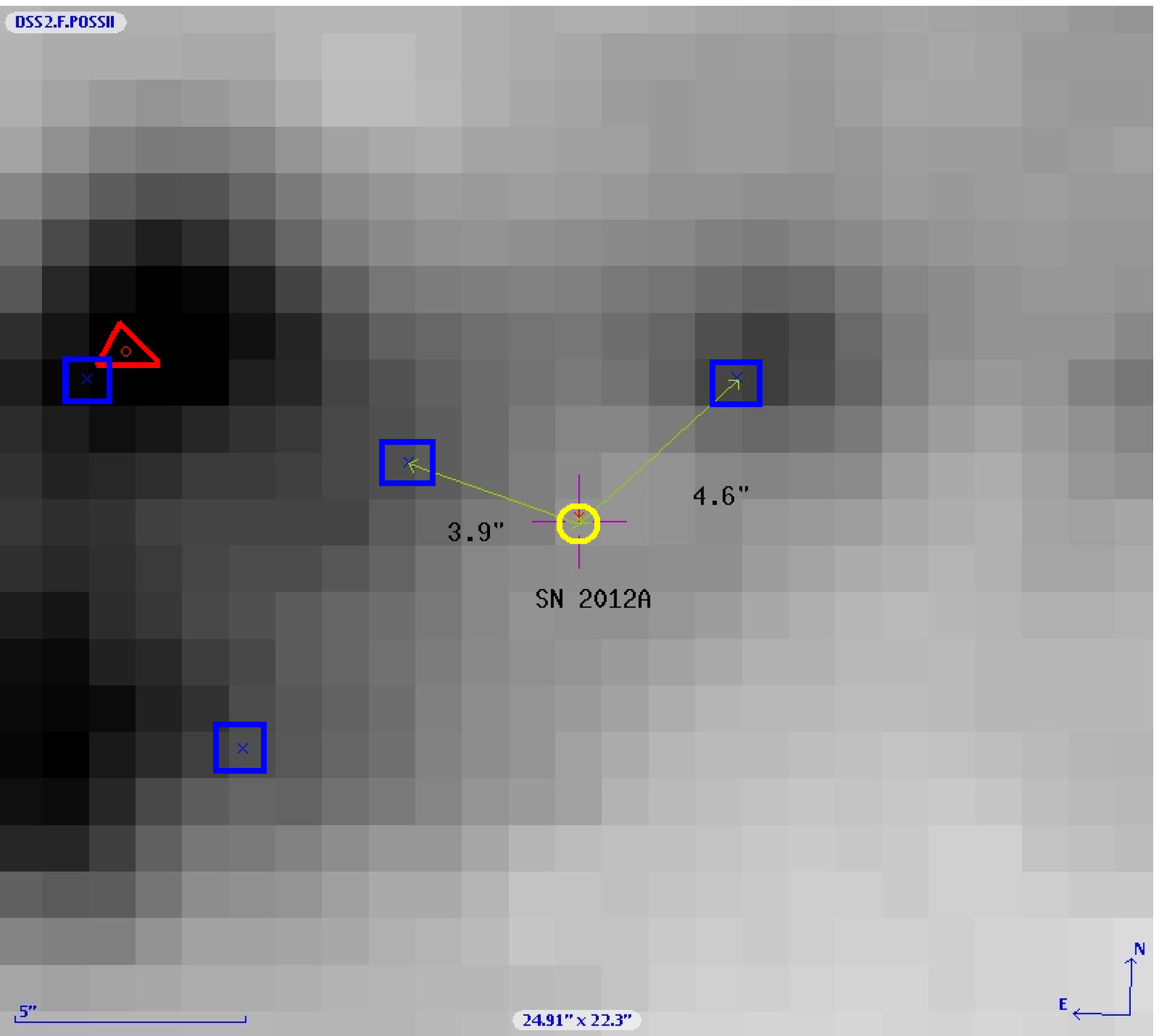}%
\caption{{\bf Left Panel:} Identification chart of the field of SN 2012A in NGC
 3239. The image is
 about 13$'\times13'$ taken in $B$-band with the 104-cm ST at ARIES, Nainital.
 The SN location is marked along with the secondary stars used for calibration.
 North is up and East is to the left. {\bf Right Panel:}
 24\mbox{.$^{''}$}91$\times$22\mbox{.$^{''}$}3 region around SN 2012A in pre-SN
 DSS image. The SN location is
 at the center and marked with the yellow circle. The red triangle is an
 extended source, whereas the blue rectangles are point sources tabulated in
 2MASS catalog. The angular separation between SN and two nearby sources are
 respectively 3\mbox{.$^{''}$}9 and 4\mbox{.$^{''}$}6. These are the prime sources of
 contamination of SN flux.}
\label{fig:SN2012A_close}
\end{figure*}

 The ground based optical photometric observations were carried out at
 the 104-cm Sampurnanand Telescope (ST) using Johnson $UBV$ and
 Cousins $RI$ filters, and also from the 130-cm Devasthal Fast Optical
 Telescope (DFOT) using $BVR$ filters. Photometric observations
 were conducted at 20 epochs between 14d and 350d. The field of SN
 2012A is calibrated using \cite[Landolt (1992)]{Landolt1992} standard stars of
 the fields of SA98. Left
 panel of Fig. \ref{fig:SN2012A_close} shows the SN position and
 location of 7 local standards in the field. The detailed methodology
 for data analysis is discussed elsewhere (\cite[Roy et
   al. 2011a,b]{Roy2011}).  A closer view of the transient location
 obtained from a pre-SN DSS image is presented in the right panel of
 Fig. \ref{fig:SN2012A_close}. Several star-forming knots, cataloged
 in the 2MASS survey and marked around the SN position, are prime
 sources of contamination of SN flux. In order to calculate the
 correct SN flux, a `template-subtraction technique' is required, but
 due to absence of pre-SN $UBVRI$ images of this field, in the present
 work the instrumental magnitudes of the SN are derived by the profile
 (PSF) fitting method. SDSS magnitudes of the nearest star forming
 knot have been used for a rough estimation of background
 flux. Therefore the measurements are expected to represent closely
 the true SN magnitude when it is bright, i.e., in the early plateau
 phase, while in later epochs the background flux is substantial, and
 hence the estimated parameters will be marginally affected.
 
 The long-slit low-resolution spectroscopy in the optical range ($0.40-0.95
 \mu{\rm m}$) was performed at eight epochs during 14d and 134d: three epochs
 from the 2-m IUCAA Girawali Observatory (IGO), one epoch from the
 3.5-m Astrophysical Research Consortium (ARC) telescope and at five
 epochs from 2-m Himalayan Chandra Telescope (HCT). We also report on
 the high resolution spectroscopy from 3.5-m ARC at the early plateau
 phase of the transient. The methodology for spectroscopic data
 analysis is same as described in \cite[Roy et al. 2011b]{Roy2011}.

\begin{figure*}
\centering
\includegraphics[width=6.5cm]{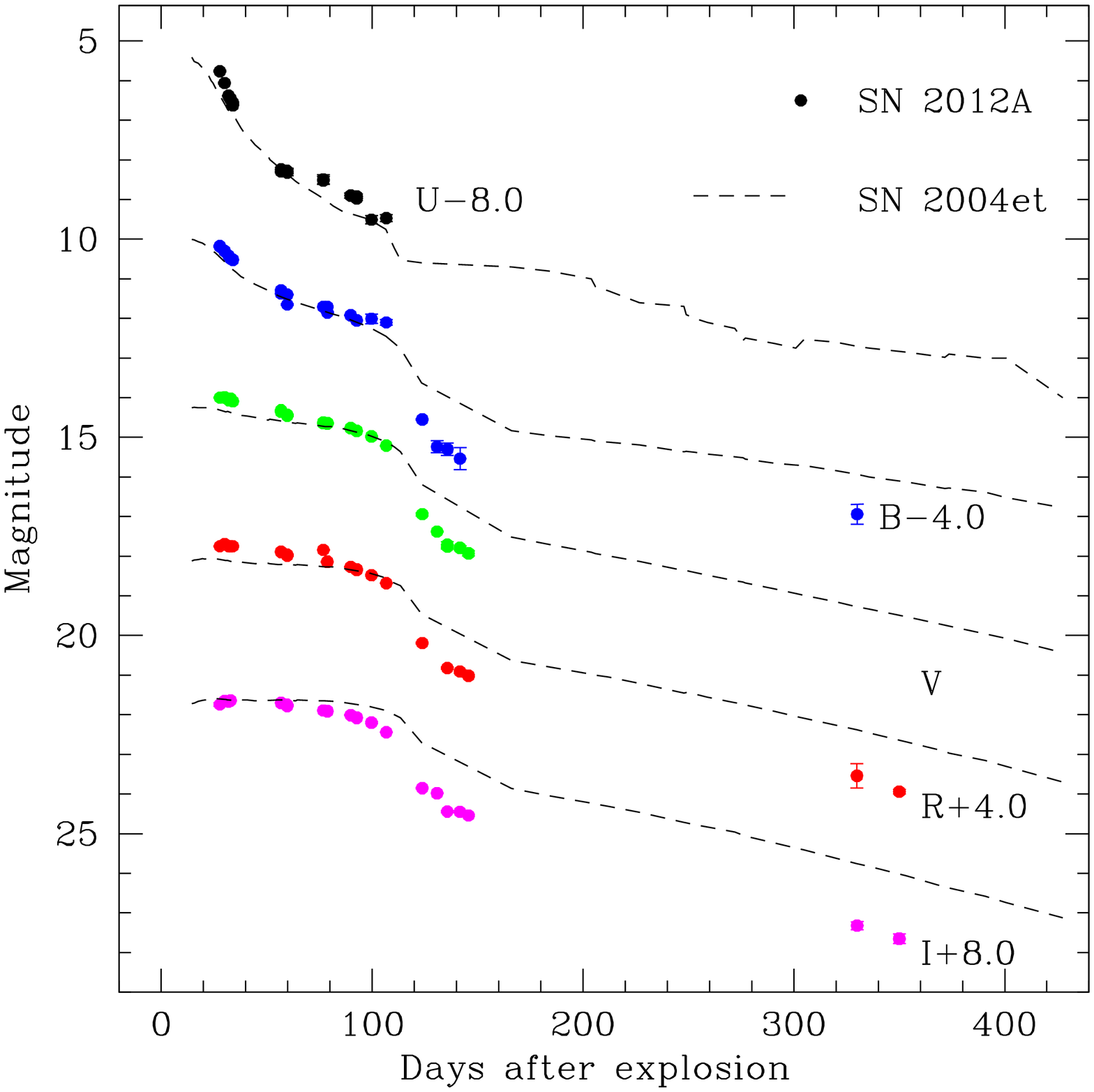}%
\hskip 1mm
\includegraphics[width=6.5cm]{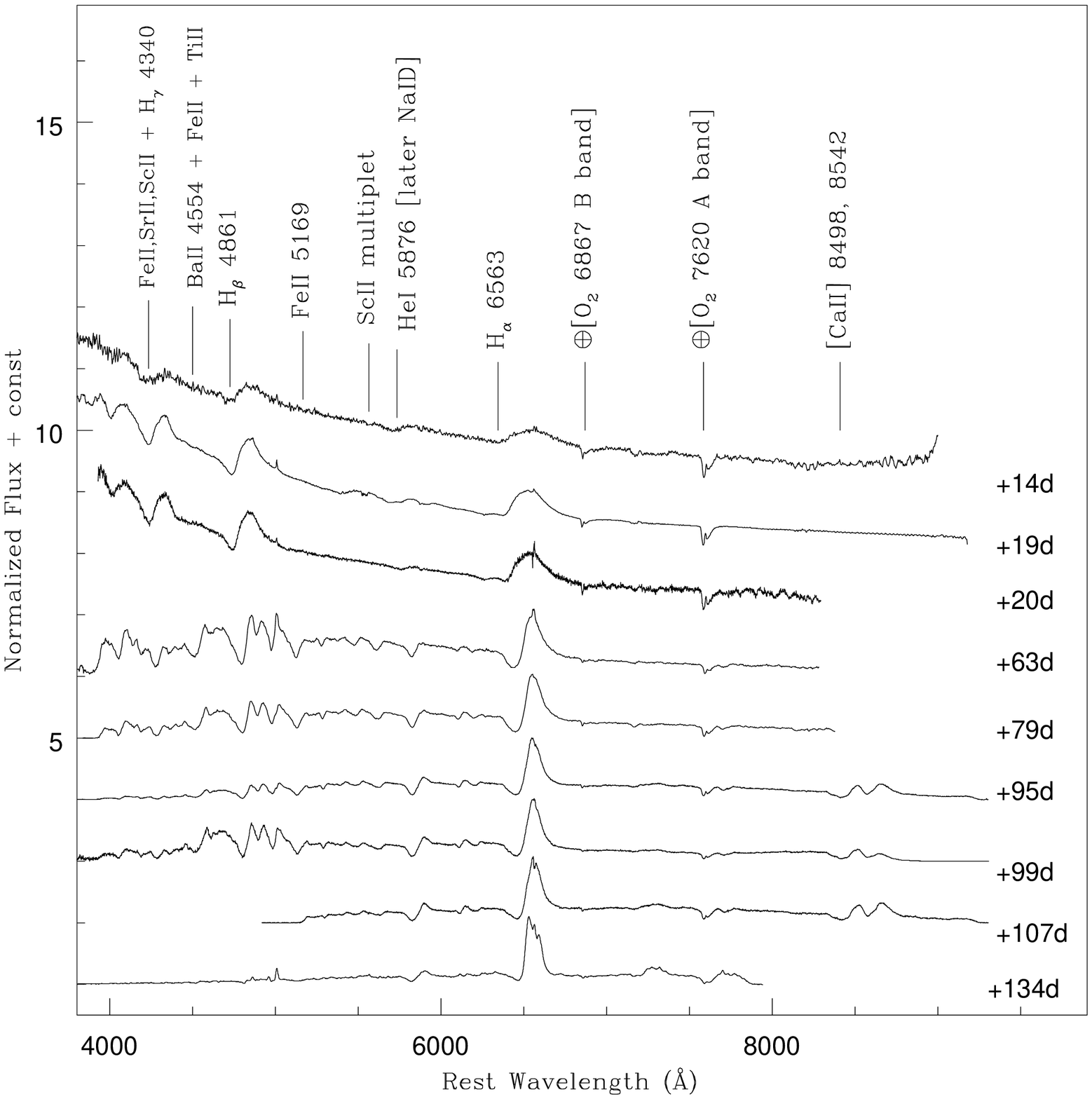}%
\caption{{\bf Left Panel:} Light curves of SN 2012A in $UBVRI$ bands. The
 light curves are shifted for clarity, while for SN 2004et they are scaled in
 magnitude and time to match with SN 2012A. {\bf Right Panel:} Doppler-corrected
 (recession velocity $\sim$752 km.s$^{-1}$) flux spectra of SN 2012A from the
 plateau (14d) to the nebular phase (134d). Prominent hydrogen and metal lines
 are marked.}
\label{fig:lc_spec}
\end{figure*}

\section{Results}
 The calibrated light curves of SN 2012A in $UBVRI$ bands are presented in the
 left panel of Fig. \ref{fig:lc_spec}, and its comparison with that of the
 normal Type IIP SN 2004et (\cite[Sahu et al. 2006]{Sahu2006}) shows that the
 plateau luminosity of both events decreases in similar fashion, though plateau
 to nebular conversion is more rapid for SN 2012A than SN 2004et.  In nebular
 phase SN 2012A becomes much fainter than SN 2004et.

\begin{figure*}
\centering
\includegraphics[width=6.5cm]{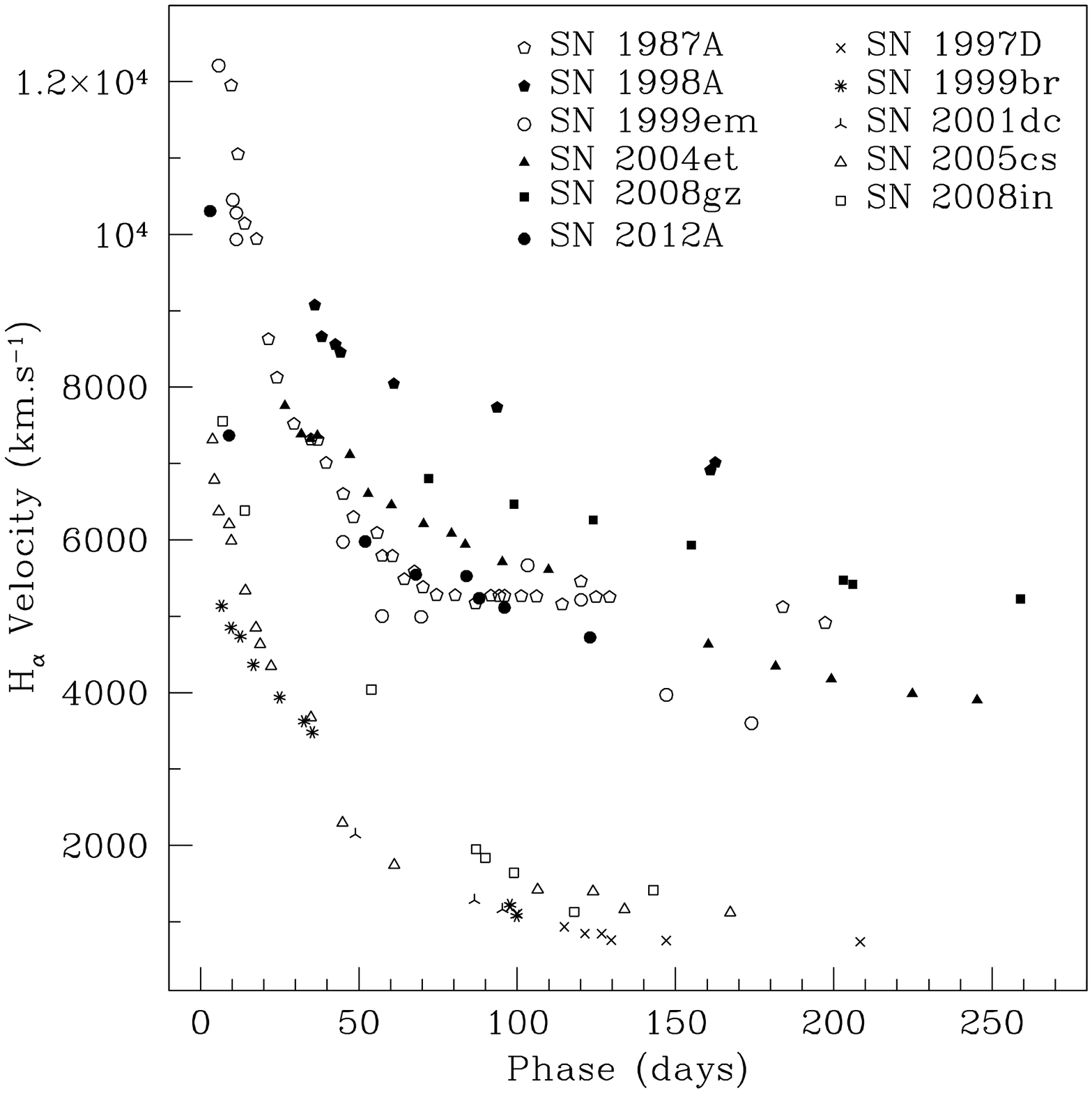}%
\hskip 1mm
\includegraphics[width=6.5cm]{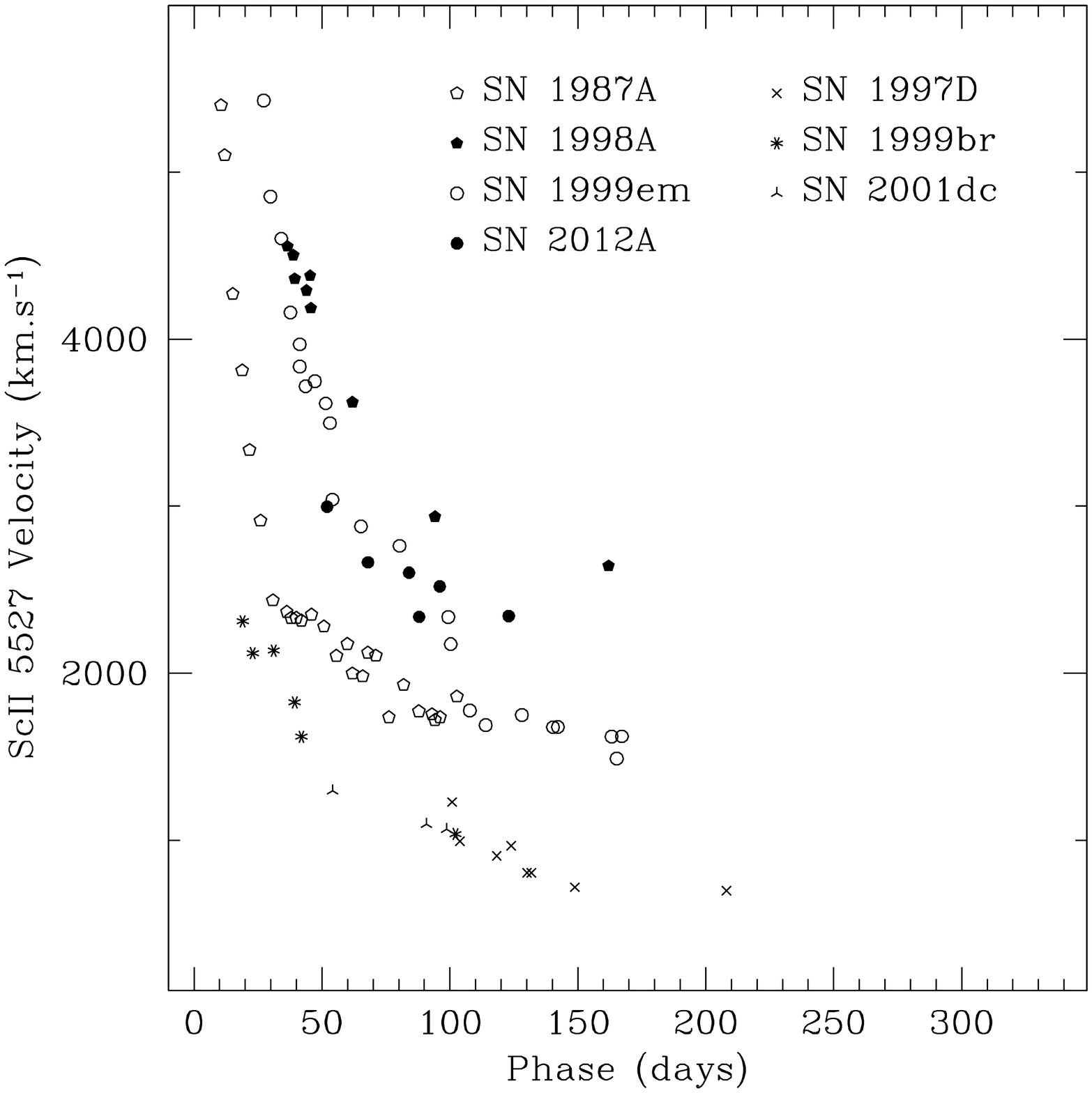}%
\caption{Comparison of velocity profile of SN 2012A with other Type IIP SNe.
 {\bf Left Panel:} H$_\alpha$ velocity, showing the velocity near outer ejecta.
 {\bf Right Panel:} ScII 5527 velocity, showing the velocity near the
 photosphere. {\it Ref:} \cite[Elmhamdi et al. (2003a)]{Elmhamdi2003},
 \cite[Pastorello et al. (2004, 2005, 2009)]{Pastorello2004_5_9},
 \cite[Roy et al. (2011a, b)]{Roy2011}, \cite[Sahu et al. (2006)]{Sahu2006}.}
\label{fig:linevel}
\end{figure*}

 The right panel of Fig. \ref{fig:lc_spec} shows the evolution of rest-frame
 spectra of SN 2012A, where the prominent lines of Hydrogen, Helium and metal
 lines along with several telluric absorption lines have been marked. The line
 identification has been done using \cite[Leonard et al (2002)]{Leonard2002}.
 The continuum dominated early spectra (14d,19d and 20d) demonstrate the high
 photospheric temperature during the initial epochs. The spectral evolution is
 similar to normal Type IIP SNe.

 Fig. \ref{fig:linevel} depicts a comparison of outer ejecta and
 photospheric velocities of SN 2012A with other Type IIP SNe.
 H$_\alpha$ profiles, representing the velocity of the outer
 ejecta, show a clear bimodal distribution: low luminosity events
 asymptotically attain a lower velocity than normal events, though
 they are much more systematic than normal Type IIP. SN 2012A is
 located near the lower edge of the velocity distribution of normal
 events. Photospheric velocity measured from the ScII 5527 line is
 almost similar in both cases, and SN 2012A preserves the
 characteristic features of normal Type IIP SNe.
  
 Galactic reddening along the line of sight, E($B-V$) =
 0.32$\pm$0.0005 mag (\cite[Schlegel et al. 1998]{Schlegel1998}) and
 the mean distance of the host is 8.6$\pm$0.5 Mpc. In our analysis of
 the high-resolution 3.5m ARC spectra, no sign of NaID at host galaxy
 velocity is found, although there is weak CaII H and K
 absorption. The upper limit on the column density ratio of NaID/CaII
 at the host galaxy velocity is very low (NaID/CaII $<$ 0.04),
 indicating very little dust grain depletion of Ca. Therefore we
 presume that extinction is mainly dominated by the Milky-Way.

From the estimated mid-plateau absolute V-band magnitude ($\sim -$15.1 mag),
 photospheric velocity ($\sim 2600$ km\,s$^{-1}$) and plateau duration ($\sim
 $110 days), the basic parameters of the progenitors have
been calculated using the prescriptions of \cite[Elmhamdi et al
  (2003b)]{Elmhamdi2003} and \cite[Litvinova and Nadezhin
  (1985)]{Litvinova1985}. The amount of synthesized $^{56}$Ni is
$\sim$ 0.012 M$_\odot$, ejected mass is $\sim$ 26 M$_\odot$, explosion
energy is $\sim 7\times10^{50}$ erg, and the radius of the
pre-SN progenitor is about 86 R$_\odot$. Assuming the mass of the
compact remnant $\sim$ 2 M$_\odot$, main sequence mass of the
progenitor can be constrained to be around 29 M$_\odot$. This is
certainly a crude estimation and more detailed modelling is essential.

\acknowledgments
 We thank George Wallerstein for providing us one of the APO echelle spectra.

\end{document}